\documentclass[twocolumn,showpacs,preprintnumbers,amsmath,amssymb, prb]{revtex4}
\usepackage{graphicx}% Include figure files
\usepackage{dcolumn}% Align table columns on decimal point
\usepackage{bm}% bold math
\begin{document}

\preprint{APS/123-QED}
%\preprint{Submitted to Phys. Rev. B on 8th Nov. 2006}

\title{Anisotropic magnetic properties of CeAg$_2$Ge$_2$
single crystals}% Force line breaks with \\
\author{A. Thamizhavel, R. Kulkarni and S. K. Dhar}
\affiliation{Department of Condensed Matter Physics and Materials
Science, Tata Institute of Fundamental Research, Homi Bhabha Road,
Colaba, Mumbai 400 005, India.}
\date{\today}% It is always \today, today,
%  but any date may be explicitly specified

\begin{abstract}
In order to investigate the anisotropic  magnetic properties of
CeAg$_2$Ge$_2$, we have successfully grown the single crystals,
for the first time, by high temperature solution growth (flux)
method.  We have performed a detailed study of
the grown single crystals by measuring their electrical resistivity,
magnetic susceptibility, magnetization, specific heat and
magnetoresistance.  A clear anisotropy and an antiferromagnetic
transition at $T_{\rm N}$~=~4.6~K have been observed in the
magnetic properties.  The magnetic entropy reaches $R$~ln~4 at 20~K indicating that the ground state and the first excited state are very closely
spaced (a quasi-quartet state). From the specific heat
measurements and crystalline electric field (CEF) analysis of the
magnetic susceptibility, we have found the level splitting
energies as 5~K and 130~K.   The magnetization measurements reveal
that the $a$-axis is the easy axis of magnetization and the
saturation moment is $\mu_{\rm s}$ = 1.6~$\mu_{\rm B}$/Ce,
corroborating the previous neutron scattering measurements on a
polycrystalline sample.

\end{abstract}

% PACS, the Physics and Astronomy
                             % Classification Scheme.

%Use showkeys class option if keyword display desired

\maketitle
\section {Introduction}
In the Ce-based intermetallic compounds, the competition
between the RKKY interaction and the Kondo effect leads to
diverse ground states.   This competition can be readily
studied in the multifarious CeT$_2$X$_2$ compounds, where T is a transition metal and X is a group IV element namely Si or Ge.  CeT$_2$X$_2$ compounds
crystallize in the well known ThCr$_2$Si$_2$ type tetragonal crystal
structure and exhibit a wide range of interesting magnetic
properties like heavy fermion superconductivity in
CeCu$_2$Si$_2$~\cite{Steglich}, pressure induced superconductivity
in CePd$_2$Si$_2$~\cite{Sheikin},
CeRh$_2$Si$_2$~\cite{Movshovich}, unconventional metamagnetic
transition in CeRu$_2$Si$_2$~\cite{Haen} etc. Similarly the
isostructural germanides also show interesting magnetic
properties~\cite{Abe, Raymond, Fukuhara, Grosche}.  While most of
the above series of silicides and germanides have been grown in
single crystalline form and the anisotropic magnetic properties
have been investigated, there are no reports available on  single crystalline CeAg$_2$Ge$_2$ owing to the difficulty in growing the
single crystal from a stoichiometric melt. Moreover, the
polycrystalline data are also limited. The first report on a polycrystalline
CeAg$_2$Ge$_2$ was made by Rauchschwalbe \textit{et
al.}~\cite{Rauchschwalbe} in which they have mentioned that this compound
undergoes an antiferromagnetic ordering below 8~K.   From neutron
scattering experiments an antiferromagnetic ordering temperature
of $T_{\rm N}$~=~7~K was reported by Knopp \textit{et
al.}~\cite{knopp} and Loidl \textit{et al.}~\cite{loidl}.  Furthermore, an ordered moment of 1.85~$\mu_{\rm B}$/Ce  at 1.5~K oriented along the [100] direction was estimated from the neutron scattering experiments.   From the specific heat measurements B$\ddot{\rm o}$hm \textit{et al}.,~\cite{Bohm}
have reported that CeAg$_2$Ge$_2$ orders antiferromagnetically
below 5~K and they observed a peak in the specific heat data at
350 mK when plotted as  $\delta {C/T}$ vs. $T$ ($\delta$C = $C$ $-$
$C_{\rm nuclear}$ $-$ $C_{\rm magnon}$) which they attributed to
coherent electronic quasi-particles of medium heavy mass,
coexisting with long range magnetic order. However, a recent
report by Cordruwisch \textit{et al.}~\cite{cordruwisch} have
reported a N\'{e}el temperature of 4.5~K on a polycrystalline
sample.  In view of these conflicting reports on the magnetic ordering temperature and to study the magnetic properties more precisely, we have succeeded in growing a single crystal of CeAg$_2$Ge$_2$
for the first time and investigated the anisotropic physical properties
by means of electrical resistivity, magnetic susceptibility,
magnetization, specific heat and magnetoresistance.

\section{Experiment}
CeAg$_2$Ge$_2$ single crystals were grown by self flux method. Since the
use of fourth element as flux normally introduces some inclusions
in the grown single crystals, we have grown the single crystals of
CeAg$_2$Ge$_2$ from an off-stoichiometric melt, with excess of Ag
and Ge.  The binary phase diagram of Ag and Ge shows an eutectic
at 650~$^\circ$C.  We have taken advantage of this eutectic
composition and used it as a flux for the growth of CeAg$_2$Ge$_2$
single crystal. Similar kind of binary eutectic compositions have been
successfully used as flux for the growth of several intermetallic
compounds like Au-Si binary eutectic for the crystal growth of
CeAu$_4$Si$_2$~\cite{nakashima}, Ag-Ge eutectic flux for the growth of YbAgGe,~\cite{morosan} Ni-Ge eutectic composition for the growth of several RNi$_2$Ge$_2$~\cite{budko} and Ni-B eutectic composition for the crystal growth of borocarbides~\cite{canfield}. The starting materials  with 3N-Ce, 5N-Ag and 5N-Ge were taken in the ratio 1 : 16.25 : 6.75 which includes the eutectic composition of the excess flux Ag-Ge.  The contents were placed in an alumina crucible, and subsequently sealed in an evacuated quartz ampoule.  The temperature of the furnace was raised to 1050~$^\circ$C and after homogenizing the mixture for two days, the furnace was cooled down to the eutectic temperature of the binary flux Ag-Ge over a period of 3 weeks time and then rapidly to room temperature.  The crystals were separated from the flux by
means of centrifuging.  The typical size of the crystal was
$6~\times~3~\times~0.3~$~mm$^3$, with (001) plane as the
flat plane.

The dc magnetic susceptibility and the magnetization measurements
were performed in the temperature range 1.8-300~K and in magnetic
fields up to 7~T along the principal directions using a Quantum
Design SQUID magnetometer. The temperature dependence of
electrical resistivity in the range 1.8-300~K was
measured using a home made DC electrical resistivity set up.  The
heat capacity and magnetoresistance measurements were performed using a Quantum Design PPMS instrument in the temperature range from  0.5~K to room
temperature and for fields up to 12~T.

\section{Experimental Results}
\subsection{X-ray studies}

Since the growth of the single crystals of CeAg$_2$Ge$_2$ was
performed from an off-stoichiometric starting composition we
performed powder X-ray diffraction by crushing a few small pieces
of the single crystals to confirm the phase purity of
%*********************FIGURE 1********************************
\begin{figure}[h]
\includegraphics[width=0.45\textwidth]{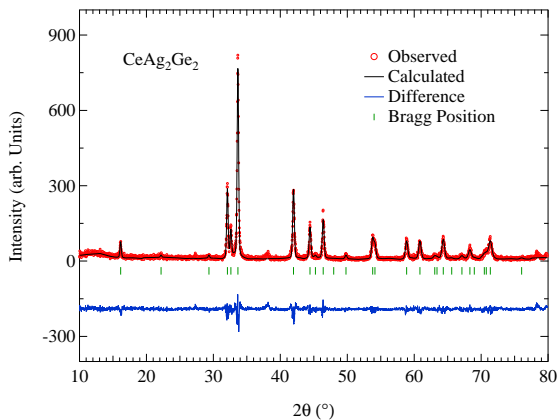}
\caption{\label{fig1}(Color online) Powder X-ray diffraction pattern recorded for
crushed single crystals of CeAg$_2$Ge$_2$ at room temperature. The
solid line through the experimental data points is the Rietveld
refinement profile calculated for the tetragonal CeAg$_2$Ge$_2$. }
\end{figure}
%****************************************************************
CeAg$_2$Ge$_2$. The powder X-ray pattern together with the
Rietveld refinement are shown in Fig.~\ref{fig1}.  The X-ray
pattern clearly reveals that the grown single crystals are single
phase and no detectable traces of impurity phases are seen. From
the Rietveld refinement the ThCr$_2$Si$_2$-type crystal structure
of CeAg$_2$Ge$_2$ is confirmed and the lattice constants were
estimated to be $a = 4.301(8)~{\rm \AA}$ and $c= 10.973(7)~{\rm
\AA}$.  We have also performed the energy dispersive X-ray
analysis (EDAX) and confirmed the stoichiometry of CeAg$_2$Ge$_2$ single crystals.  The crystals were then oriented along the principal directions, namely [100] and [001] directions, by means of the Laue back reflection
method.  Well defined Laue diffraction spots, together with the
tetragonal symmetry pattern, indicated the good quality of the
single crystals.  The crystals were cut along the principal
direction using a spark erosion cutting machine for the
anisotropic physical property measurements.

\subsection{Electrical resistivity}
The dc electrical resistivity of CeAg$_2$Ge$_2$ in the temperature range from
%*********************FIGURE 2********************************
\begin{figure}[h]
\includegraphics[width=0.45\textwidth]{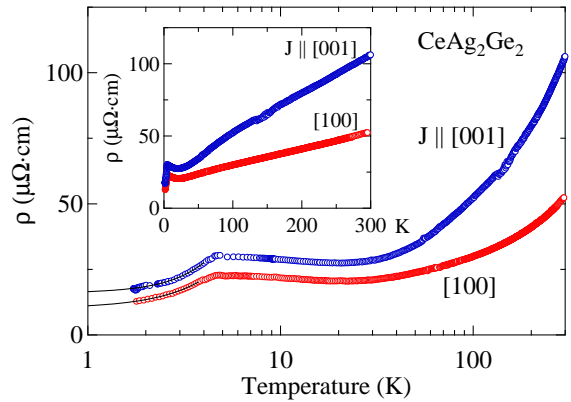}
\caption{\label{fig2}(Color online) Logarithmic temperature dependence of the dc electrical resistivity of  CeAg$_2$Ge$_2$ for J~$\parallel$~[100] and [001]. The inset shows the linear temperature dependence of the electrical resistivity.  The solid lines are least square fitting toa power law relation.}
\end{figure}
%****************************************************************
1.8 to 300~K is shown in Fig.~\ref{fig2}.  The resistivity was measured for the current direction parallel to [100] and [001].  The electrical resistivity is anisotropic reflecting the tetragonal symmetry of the crystal structure.  As it can be seen from the Fig.~\ref{fig2} the absolute value of electrical resistivity at 295~K is 52~$\mu\Omega\cdot$cm and 106~$\mu\Omega\cdot$cm, respectively for $J~\parallel$~[100] and [001] and at 1.8~K is 12~$\mu\Omega\cdot$cm and 18~$\mu\Omega\cdot$cm for $J~\parallel$~[100] and [001], respectively.  At high temperatures the scattering is phonon dominated and the resistivity decreases linearly with decreasing temperature typical of a metallic sample. The electrical resistivity shows a shallow minimum around 20~K and then increases with decrease in temperature up to 4.6~K.  This increase in the electrical resistivity at low temperature can be attributed to short range antiferromagnetic order and/or the presence of weak Kondo-type interaction. It may be mentioned here that the corresponding silicide, CeAg$_2$Si$_2$ has been reported to be a dense Kondo lattice antiferromagnet~\cite{Garde}. With further decrease in temperature below 4.6~K, the resistivity changes its slope and drops due to the reduction in spin-disorder scattering caused by the antiferromagnetic ordering of the magnetic moments.  In the limited temperature range from 1.8 - 4.0~K the resisitivity follows a power law relation $\rho = \rho_{0} + AT^{n}$ with $\rho_0$ = 10.15~$\mu \Omega \cdot$cm, $A= 0.99~\mu\Omega\cdot$cm/K$^{1.75}$ and $n = 1.75$  and $\rho_0$ = 15.97~$\mu \Omega \cdot$cm, $A= 0.53~\mu\Omega\cdot$cm/K$^{2.21}$ and $n = 2.21$ for the currents along $J~\parallel~$[100] and [001] directions, respectively.  Here the exponent $n$ is close to 2 which can in principle be explained on the basis of electron-electron scattering. Since our data do not extend for $T << T_{\rm N}$ we have not attempted to fit our data using sophisticated model of spin-waves.  

\subsection{Magnetic susceptibility and magnetization} 
The temperature dependence of magnetic susceptibility in the temperature range from 1.8 to 300~K measured in a field of 1~kOe along the two principal 
%*********************FIGURE 3********************************
\begin{figure}[h]
\includegraphics[width=0.45\textwidth]{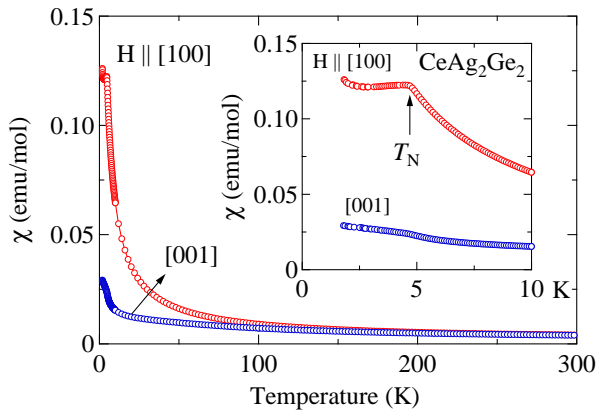}
\caption{\label{fig3}(Color online)Temperature dependence of magnetic susceptibility $\chi$ from 1.8 -- 300~K.  The inset shows the low temperature magnetic susceptibility.}
\end{figure}
%****************************************************************
directions viz., $H$ parallel to [100] and [001] is shown in Fig.~\ref{fig3}.  The antiferromagnetic ordering at $T_{\rm N}$ = 4.6~K is clearly seen as indicated by the arrow.  The susceptibility below $T_{\rm N}$ remains almost $T$-independent and at the lowest temperature measured there is a small rise in the susceptibility, indicating that the antiferromagnetism observed in CeAg$_2$Ge$_2$ is not a simple two sublattice antiferromagnetism. For example, canted antiferromagnetism may show a weak residual ferromagnetic magnetization in the N\'{e}el state.  Also, this type of temperature independent susceptibility at low temperature may be attributed to the crystalline electric field (CEF) effect.  The inverse magnetic susceptibility of CeAg$_2$Ge$_2$  does not obey the simple Curie-Weiss law (not shown here), on the other hand, it can be very well fitted to a modified Curie-Weiss law which is given by $\chi=\chi_0 + \frac{C}{T-\theta_p}$,
where $\chi_0$ is the temperature-independent part of the magnetic susceptibility and $C$ is the Curie constant.  The main contributions to $\chi_0$ includes the core-electron diamagnetism, and the susceptibility of the conduction electrons. The details of the inverse magnetic susceptibility is discussed later in the discussion part. For an effective magnetic moment of 2.54~$\mu_{\rm B}$/Ce we have estimated the $\theta_{\rm p}$ values as -7.2~K and -42~K for $H~\parallel$~[100] and [001], respectively.  

\begin{figure}[h]
\includegraphics[width=0.45\textwidth]{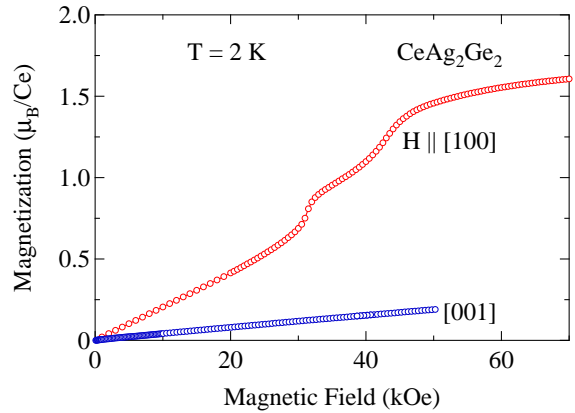}
\caption{\label{fig4}(Color online) Isothermal magnetization curves of CeAg$_2$Ge$_2$ measured at $T=2$~K along the two principal directions. }
\end{figure}

\begin{figure}[h]
\includegraphics[width=0.45\textwidth]{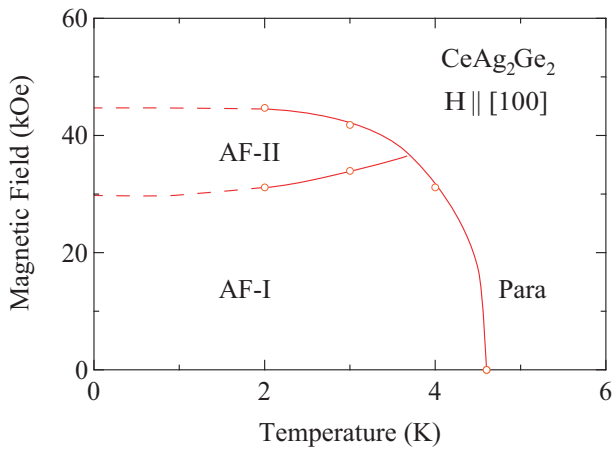}
\caption{\label{fig5}(Color online) Magnetic phase diagram of CeAg$_2$Ge$_2$ for $H~\parallel$~[100].}
\end{figure}

The field dependence of isothermal magnetization at $T$~=~2~K measured in a SQUID magnetometer up to a field of 70~kOe is shown in Fig.~\ref{fig4}.  The 
%*********************FIGURE 4********************************
%****************************************************************
magnetization curves show large uniaxial magnetocrystalline anisotropy.  The magnetization for $H~\parallel$~[100] is linear for low fields and shows  metamagnetic transitions at  critical fields $H_{\rm m1}$ = 31~kOe and at $H_{\rm m2}$ = 44.7~kOe and nearly saturates at 70~kOe with a saturation moment $\mu_{\rm s}$ = 1.6~$\mu_{\rm B}$/Ce, this indicates that [100]-axis is the easy axis of magnetization. Here, the saturation moment is smaller than the free ion value of 2.1~$\mu_{\rm B}$/Ce which could be attributed to the crystal field effects. However, one can achieve the saturation value at high applied magnetic fields. On the other hand, the magnetization for $H~\parallel$~[001] is very small and varies linearly with field reaching a value of 0.32~$\mu_{\rm B}$/Ce at 50~kOe, indicating a hard axis of magnetization.  We have also performed the isothermal magnetization at 3~K, 4~K, 5~K and 10~K for $H~\parallel$~[100].  From the differential plots of the isothermal magnetization measurements, we have constructed the magnetic phase diagram as shown in Fig.~\ref{fig5}. The two metamagnetic transitions are clearly seen for 2~K and 3~K magnetization curves; however at 3~K only one metamagnetic transition is seen.  For temperatures above the magnetic ordering temperature the magnetization curves did not show any metamagnetic behaviour and the magnetization curves were linear indicating a paramagnetic state.   
%*********************FIGURE 5********************************
%****************************************************************

\subsection{Specific heat}
\begin{figure}[h]
\includegraphics[width=0.45\textwidth]{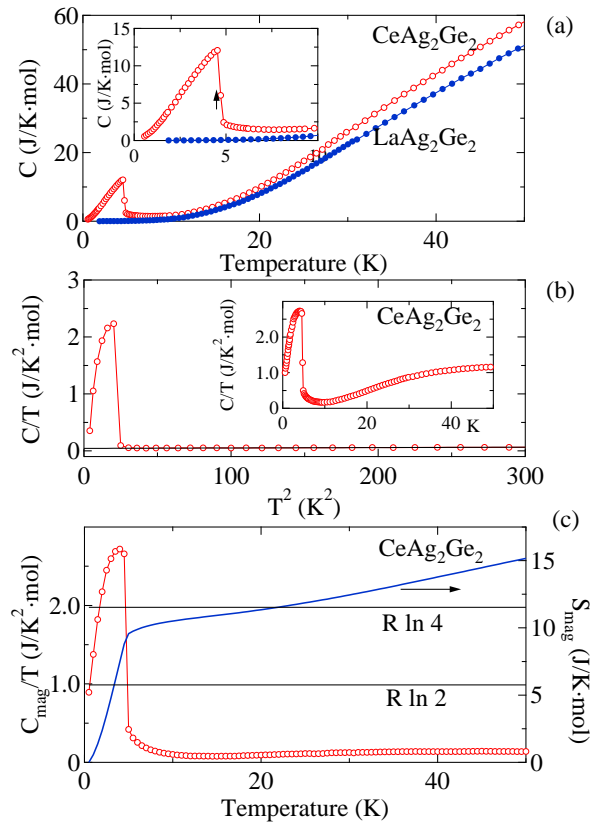}
\caption{\label{fig6}(Color online)(a) Temperature dependence of the specific heat of CeAg$_2$Ge$_2$  and LaAg$_2$Ge$_2$,  the inset shows the low temperature part, (b) specific heat of CeAg$_2$Ge$_2$ in the form of $C/T$ vs. $T^2$ after subtracting the schottky and the 4$f$ contribution, the solid line shows the extrapolation of the high temperature specific heat to 0~K,  inset shows $C/T$ versus $T$ and (c) magnetic specific heat $C_{\rm mag}$ in the form of $C_{\rm mag}/T$ vs. $T$ together with the magnetic entropy $S_{\rm mag}$.}
\end{figure}

Figure~\ref{fig6}(a) shows the temperature dependence of the specific heat of  
%*********************FIGURE 6********************************
%****************************************************************
single crystalline CeAg$_2$Ge$_2$ together with the specific heat of a polycrytalline reference sample LaAg$_2$Ge$_2$. The low temperature data ($\sim$~1.5 to $\sim$~10~K) of LaAg$_2$Ge$_2$ have been fitted to the expression $C = \gamma T + \beta T^3$ where $\gamma$ is the electronic contribution and $\beta$ is the phonon contribution to the heat capacity.  The $\gamma$ and $\beta$ values thus obtained are estimated to be 2.8~mJ/K$^2$$\cdot$mol and 0.59~mJ/K$^4$$\cdot$mol, respectively.   The inset of Fig.~\ref{fig6}(a) shows the low temperature part of the specific heat and the antiferromagnetic ordering is manifested by the clear jump in the specific heat at $T_{\rm N}$ = 4.6~K as indicated by the arrow. The inset of Fig.~\ref{fig6}(b) shows the specific heat in the form of $C/T$ versus $T$.  Just below the magnetic ordering the specific heat shows a broad peak in the $C/T$ versus $T$ curve which presumably indicates the presence of low lying crystal field levels. Assuming the lattice heat capacity of CeAg$_2$Ge$_2$ is the same as that of LaAg$_2$Ge$_2$, the $4f$-derived contribution to the heat capacity $C_{\rm mag}$ was obtained by subtracting the specific heat of LaAg$_2$Ge$_2$ from the total specific heat of CeAg$_2$Ge$_2$.  Figure~\ref{fig6}(c) shows  $C_{\rm mag}/T$ versus $T$ together with the entropy $S_{\rm mag}$ which is obtained by integrating $C_{\rm mag}/T$.  As it can be seen from the figure, the entropy of CeAg$_2$Ge$_2$ is very high at the magnetic ordering temperature and reaches $R$~ln~4 near 20~K.  In tetragonal symmetry, the degenerate six fold levels of the ground-state multiplet of Ce$^{3+}$ split into three doublets and $\Delta_1$ and $\Delta_2$ are the excitation energies of the first and second excited states, respectively.  Since the entropy change  reaches $R$~ln~4, not too far above $T_{\rm N}$, one can come to a conclusion that the ground state and the first excited state are very closely spaced or nearly degenerate.  This finding clearly corroborates the earlier neutron scattering results by Loidl~\textit{et al.}~\cite{loidl} in which they could observe only one crystal field transition at 11~meV and concluded that the ground state is almost degenerate with the first excited state.  

From the crystalline electric field analysis of the magnetic susceptibility data, to be discussed later, we found that the energies of the exicted states $\Delta_1$ and $\Delta_2$ as 5~K and 130~K, respectively.  Due to the very small splitting energy between the ground state and the first excited state the estimation of the Sommerfeld coefficient $\gamma$, by the usual method, from the low temperature data will lead to ambiguity.  Hence we estimated the $\gamma$ value from the high temperature data in the paramagnetic region above the magnetic ordering after subtracting the Schottky contribution and linearly extrapolating the $C/T$ versus $T^2$ behaviour to T~=~0~K and is shown in  Fig.~\ref{fig6}(b).  The $\gamma$ value thus estimated is 45~mJ/K$^2\cdot$mol.

\subsection{Magnetoresistance}

We have also studied the effect of magnetic field on the resistivity of CeAg$_2$Ge$_2$.  The magnetic field did not have any appreciable effect on the resistivity for the field perpendicular to the easy axis direction.  On the other hand, when the field was applied parallel to the easy axis direction, we found that the resistivity gradually decreased with increasing field.  In
%*********************FIGURE 8********************************
\begin{figure}[h]
\includegraphics[width=0.45\textwidth]{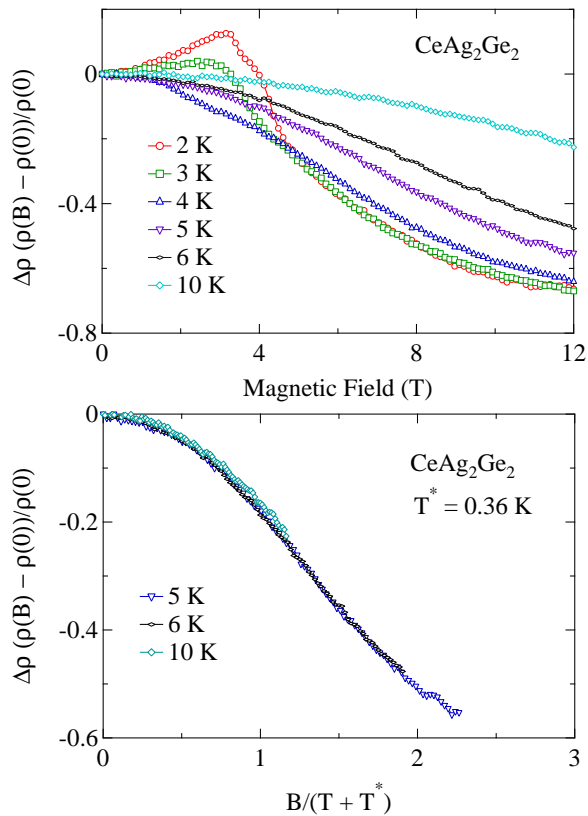}
\caption{\label{fig8}(Color online) (a)Isothermal normalized magnetoresistance for CeAg$_2$Ge$_2$ as  function of applied magnetic field for $J~\parallel~$[100] and $H~\parallel~$ [100]. (b) Normalized magnetoresistivity, in the paramagnetic state, plotted as a function of $B/(T+T^*)$.}
\end{figure}
%****************************************************************
 Fig.~\ref{fig8} we have plotted the normalized magnetoresistance $\Delta \rho/\rho_0$ = [$\rho(B)$ -- $\rho(B = 0)$]/$\rho(B = 0)$ as a function of applied magnetic field at various fixed temperatures.  With increasing magnetic field for $H~\parallel$~[100], $\Delta \rho/\rho_0$ at 2~K initially increases in the positive direction and then turns negative at higher magnetic fields, giving rise to a maximum at 3.1~T.  This field value coincides with the metamagnetic transition observed in the magnetization measurement at T = 2~K.  Such a behaviour of the magnetoresistance is qualitatively consistent with the theoretical calculation given by Yamada and Takada~\cite{yamada}.  In the antiferromagnetic state ($H < H_m$), the magnetic moment fluctuation in one magnetic sublattice is enhanced by the field while, in the field induced ferromagnetic state ($H > H_m$), the fluctuation is suppressed by the field.  The change in the fluctuation is reflected in the magnetoresistance.  With the increase in the temperature, the peak in the  magnetoresistance moves toward lower fields and decreases, finally disappears for temperatures above $T_{\rm N}$.  In the paramagnetic region the negative magnetoresistance is due to the freezing out of the spin-flip scattering by the magnetic field.  The normalized magnetoresistance for $T > T_{\rm N}$ can be mapped onto a single curve using the scaling relation $\Delta \rho/\rho(0)$ =  $f[B/(T + T^*)]$ derived by Schlottmann\cite{schlottmann} within the Bethe-ansatz approach, as shown in Fig.~\ref{fig8}(b).  Here $T^*$ is the characteristic temperature which is an approximate measure of the Kondo temperature $T_{\rm K}$ and is estimated to be 0.36~K.  This indicates that the Kondo effect is very weak in CeAg$_2$Ge$_2$, which substantiates our earlier prediction from the zero field resistivity data.    

\section{Discussion}
From the results of the electrical resistivity, susceptibility and specific heat measurements it can be clearly seen  that CeAg$_2$Ge$_2$ undergoes an antiferromagnetic ordering at 4.6~K with the easy axis of magnetization as [100].  The magnetization at 70~kOe reaches 1.6~$\mu_{\rm B}$/Ce thus corroborating the earlier neutron scattering experiment~\cite{loidl} on a polycrystalline sample of CeAg$_2$Ge$_2$.  The electrical resistivity at high temperature shows a typical metallic behaviour and at sufficiently low temperature it shows a weak minimum before ordering magnetically.  This behaviour is quite different from what one has observed in the CeCu$_2$Ge$_2$ which is similar to CeAg$_2$Ge$_2$ both structurally and magnetically, although the antiferromagnetic ordering temperature is nearly equal ($T_{\rm N}$ = 4.1~K for CeCu$_2$Ge$_2$ and $T_{\rm N}$ = 4.6~K for CeAg$_2$Ge$_2$).  The logarithmic temperature dependence of electrical resistivity in CeCu$_2$Ge$_2$ exhibit a double peak structure which is presumably attributed to the combined influence of the Kondo and crystalline electric field (CEF) effects.  The Kondo temperature of CeCu$_2$Ge$_2$ was estimated to be about 6~K~\cite{knopp2}, whereas for CeAg$_2$Ge$_2$ the Kondo temperature is very small. Since the unit cell volume of CeCu$_2$Ge$_2$ is smaller (V~$\approx$~178~{\rm \AA}$^3$), a  larger value of the Kondo coupling constant $J_{sf}$ is expected and hence the Kondo interaction dominates  in CeCu$_2$Ge$_2$ compared to that in CeAg$_2$Ge$_2$(V~$\approx$~203~{\rm \AA}$^3$).  In CeCu$_2$Ge$_2$ superconductivity occurs when the unit cell volume attains a favourable value of 168~$\pm$~3~\AA$^3$.  This is achieved with an external pressure of 7~GPa.  Considering this fact the unit cell volume of CeAg$_2$Ge$_2$ is quite large and one would require a very high pressure to reduce the unit cell volume to nearly 168~$\pm$~3 \AA$^3$ for probable observation of superconductivity.  Based on this it can be said that CeAg$_2$Ge$_2$ lies on the left hand side of the Doniach phase diagram in which the RKKY energy scale is dominant and Kondo interaction is weak.

The heat capacity measurement of CeAg$_2$Ge$_2$ single crystal clearly reveals the presence of low lying crystal field levels with a very small separation between the ground state and the first excited state indicating that the ground state is a quasi-quartet state instead of a doublet which is usually observed for a tetragonal site symmetry.  In order to further analyze the crystal field
%*********************FIGURE 9********************************
\begin{figure}[h]
\includegraphics[width=0.45\textwidth]{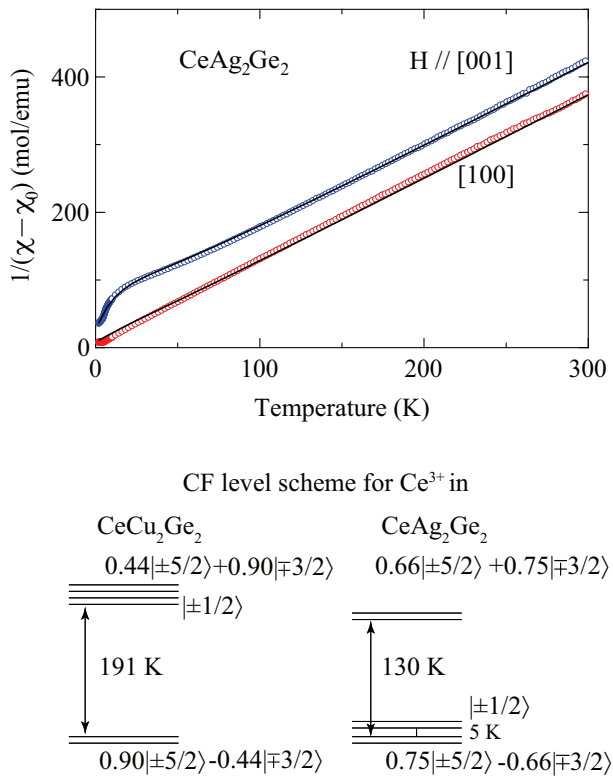}
\caption{\label{fig9}(Color online) Temperature dependence of the inverse magnetic susceptibility in CeAg$_2$Ge$_2$.  Solid lines are fitting to the CEF level scheme with a molecular field term.  The bottom figure shows the crystal-field level scheme of Ce$^{3+}$ in the CeCu$_2$Ge$_2$ (taken from Ref.21) and CeAg$_2$Ge$_2$.}
\end{figure}
%****************************************************************
levels and to understand the present anisotropy in the magnetic susceptibility we have performed the CEF analysis on the susceptibility data.   For the purpose of CEF analysis in Fig.~\ref{fig9} we have plotted the experimental results on susceptibility in the form of $1/(\chi - \chi_0)$, where $\chi_0$ was determined as 1.33~$\times$~10$^{-3}$ and 1.41~$\times$~10$^{-3}$ emu/mol for $H~\parallel$~[001] and [100], respectively, so an effective magnetic moment of 2.54~$\mu_{\rm B}$/Ce is obtained above 100~K.  Similar kind of treatment has been made for CePt$_3$Si~\cite{takeuchi} while performing the CEF analysis. The Ce atoms in CeAg$_2$Ge$_2$ occupy the $2a$ Wyckoff position with the  point symmetry 4/mmm ($D_{\rm 4h}$)  and hence possess the tetragonal site symmetry.  The CEF hamiltonian for a tetragonal site symmetry is given by,
\begin{equation}
\label{eq:HCEF}
\mathcal{H}_{\rm CEF} = B_{2}^{0}O_{2}^{0} + B_{4}^{0}O_{4}^{0} + B_{4}^{4}O_{4}^{4} + B_{6}^{0}O_{6}^{0} + B_{6}^{4}O_{6}^{4},
\end{equation}
where $B_{\ell}^{m}$ and $O_{\ell}^{m}$ are the CEF parameters and the Stevens operators, respectively~\cite{Stevens,Hutchings}.  For Ce atom, the 6th order terms $O_{6}^{0}$ and $O_{6}^{4}$ vanishes and hence CEF Hamiltonian reduces to,  
\begin{equation}
\label{eq:HCEF}
\mathcal{H}_{\rm CEF} = B_{2}^{0}O_{2}^{0} + B_{4}^{0}O_{4}^{0} + B_{4}^{4}O_{4}^{4},
\end{equation}
The magnetic susceptibility including the molecular field contribution $\lambda$ is given
by
\begin{equation}
\label{eq:chi}
\chi^{-1} = \chi_{\rm CEF}^{-1} - \lambda,
\end{equation}
where $\chi_{\rm CEF}$ is the CEF susceptibility.  Diagonalization of the CEF Hamiltonian gives us the eignvalues and eigenfunctions. For Ce$^{3+}$ $J=5/2$ wave function splits into three doublets, $\Gamma_{7}^{(1)} = a\left|\pm5/2\right\rangle +b\left|\mp3/2\right\rangle$, $\Gamma_{7}^{(2)} = a\left|\pm3/2\right\rangle -b\left|\mp5/2\right\rangle$ and $\Gamma_6 =  \left|\pm1/2\right\rangle$, where $a$ and $b$ are mixing parameters with the condition $a^2 + b^2=1$.  The CEF parameters were estimated from the fits to the magnetic susceptibility. Solid lines in Fig.~\ref{fig9} show the least square fitting to Eqn.~\ref{eq:chi}, the CEF parameters thus obtained are listed in Table~\ref{tab:table1}.  The corresponding crystal field level scheme together with that of CeCu$_2$Ge$_2$ is shown in the bottom part of Fig.~\ref{fig9}.  The crystal field level scheme for CeCu$_2$Ge$_2$ is taken from Ref.~21.  The ground state of CeAg$_2$Ge$_2$ shows a mixing of $\left|\mp3/2\right\rangle$ and $\left|\pm5/2\right\rangle$ wave functions.
%***************************** Table 2 **********************************
\begin{table*}
\caption{\label{tab:table1} CEF parameters, energy level schemes
and the corresponding wave functions for CeAg$_2$Ge$_2$.}
\begin{ruledtabular}
\begin{tabular}{ccccccc}
CEF parameters    \\ \hline
             & $B_{2}^{0}$~(K) & $B_{4}^{0}$~(K) & $B_{4}^{4}$~(K) & $\lambda_{i}$~(emu/mol)$^{-1}$ \\
& $2.24$ & $-0.19$ & $2.40$ & $\lambda_{[100]}$ = $-7.2$ \\
&      &         &   & $\lambda_{[001]}$ = $-26$  \\ \hline
           energy levels and wave functions     \\ \hline
$E$(K) & $\mid+5/2\rangle$ & $\mid+3/2\rangle$ & $\mid+1/2\rangle$ & $\mid-1/2\rangle$ & $\mid-3/2\rangle$ & $\mid-5/2\rangle$ \\
130 & $-0.656$ & 0 & 0 & 0 & $-0.754$ & 0 \\
130 & 0 & $-0.754$ & 0 & 0 & 0 & $-0.656$ \\
5  & 0 & 0 & 0 & 1 & 0 & 0 \\
5  & 0 & 0 & 1 & 0 & 0 & 0 \\
0 & 0 & $-0.656$ & 0 & 0 & 0 & $0.754$ \\
0 & $0.754$ & 0 & 0 & 0 & $-0.6565$ & 0 \\
\end{tabular}
\end{ruledtabular}
\end{table*}
%***********************************************************************
From Fig.~\ref{fig9} it is obvious that the present set of CEF parameters gives a good fit to the experimental data there by explaining the anisotropy in the magnetic susceptibility. The CEF parameters have resulted in the first and  second excited states at $\Delta_1$~=~5~K and $\Delta_2$~=~130~K.  It is interesting to note here that the CEF level scheme of CeCu$_2$Ge$_2$ is qualitatively opposite to the present case where the ground state is a doublet and the first and second excited states are nearly degenerate.  This can be explained on the basis of the sign of the $B_{2}^{0}$ parameter.  For CeCu$_2$Ge$_2$ it is negative while it is positive for CeAg$_2$Ge$_2$.  This change in sign of  $B_{2}^{0}$ suggests that the CEF potential in CeT$_2$Ge$_2$ is largely dependent on the hybridization  between localized $f$-electron states and the conduction-electron bands.

\section{conclusion}

Single crystals of CeAg$_2$Ge$_2$ have been grown for the first time, by flux method by using a Ag-Ge binary eutectic composition as flux. The antiferromagnetic ordering temperature $T_{\rm N}$ = 4.6~K is clearly manifested by the resistivity, heat capacity and susceptibility measurements.   Thus the ambiguity about the magnetic ordering temperature of this compound, reflected in the conflicting reports earlier in the literature, has been removed. A large anisotropy in the electrical resistivity, magnetic susceptibility and magnetization is observed.  The susceptibility and magnetization clearly reveals that [100]-axis as the easy axis of magnetization with a moment of 1.6~$\mu_{\rm B}$/Ce at 70~kOe.  Metamagnetic transitions have been observed at  the critical fields, $H_{\rm m1}$ = 31~kOe and at $H_{\rm m2}$ = 44.7~kOe. The heat capacity and the susceptibility data clearly support the closely spaced ground and first excited states, which have been analyzed by the CEF calculations.

\end{document}